\documentclass[%
 aip,
 amsmath,amssymb,
reprint, onecolumn 
]{revtex4-1}

\usepackage{graphicx}
\usepackage{dcolumn}
\usepackage{braket}
\usepackage{bm}

\usepackage[utf8]{inputenc}
\usepackage[T1]{fontenc}
\usepackage{mathptmx}
\usepackage{xcolor,colortbl}
\usepackage{etoolbox}
\newtheorem{theorem}{Theorem}

\newtheorem{proposition}{Proposition}
\newtheorem{observation}{Observation}
\newtheorem{example}{Example}
\newtheorem{remark}{Remark}
\newtheorem{lemma}{Lemma}
\newtheorem{result}{Result}
\usepackage{amsmath}
\usepackage[colorlinks=true,
            linkcolor=blue,
            citecolor=blue,
            urlcolor=blue]{hyperref}

\makeatletter
\def\@email#1#2{%
 \endgroup
 \patchcmd{\titleblock@produce}
  {\frontmatter@RRAPformat}
  {\frontmatter@RRAPformat{\produce@RRAP{*#1\href{mailto:#2}{#2}}}\frontmatter@RRAPformat}
  {}{}
}%
\makeatother

\begin{document}


\title{\vspace{1cm}On the characterization of partially entanglement breaking and annihilating channels}

\author{
\vspace{0.5cm}
Bivas Mallick$^{1}$\footnote{bivasqic@gmail.com},
Nirman Ganguly$^{2}$\footnote{nirmanganguly@hyderabad.bits-pilani.ac.in}, and 
A.\ S.\ Majumdar$^{1}$\footnote{archan@bose.res.in}
\\[2mm]
$^{1}$S. N. Bose National Centre for Basic Sciences, Block JD, Sector III, Salt Lake, Kolkata 700106, India
\\
$^{2}$Department of Mathematics, Birla Institute of Technology and Science Pilani, Hyderabad Campus, Hyderabad, Telangana 500078, India
}

\begin{abstract}
 Transmission of high dimensional entanglement through quantum channels is a significant area of interest in quantum information science. The certification of high dimensional entanglement is usually done through Schmidt numbers, which quantify the entanglement dimensionality of quantum states. States with high Schmidt numbers provide a larger advantage in various quantum information processing tasks compared to quantum states with low Schmidt numbers. However, the action of quantum channels may reduce the Schmidt number of transmitted states, thereby degrading their resourcefulness. Here we present a comprehensive analysis of partially entanglement breaking channels which reduce the Schmidt number of bipartite composite systems. From a resource theoretic perspective, it becomes imperative to identify channels that preserve the Schmidt number. Based on our characterization we lay down prescriptions to identify such channels which are non-resource breaking, i.e., preserve the Schmidt number. Additionally, we introduce a new class of quantum channels, termed partially entanglement annihilating channels which reduce the Schmidt number of a quantum state that is a part of a larger composite system. Finally, we study the connection between entanglement breaking, partially entanglement breaking, and partially entanglement annihilating channels.
\end{abstract}

\maketitle

\section{Introduction}

Quantum entanglement is a fundamental phenomenon in quantum theory that serves as the foundation for the rapidly evolving field of quantum technologies \cite{peres1997quantum,bohr1935can,einstein1935can,horodecki2009quantum,plenio2014introduction,nielsen2001quantum}. Entanglement occupies a privileged position in the study of quantum information processing protocols as it is a key ingredient in various tasks such as quantum teleportation \cite{bennett1993teleporting}, super dense coding \cite{bennett1992communication}, quantum cryptography \cite{ekert1991quantum,branciard2012one}, remote state preparation \cite{pati2000minimum},  secret sharing \cite{hillery1999quantum,cleve1999share,bandyopadhyay2000teleportation}. However, the most challenging aspect in this domain is to develop a well-defined framework for the detection of entanglement \cite{lewenstein2001characterization,spengler2012entanglement,shang2018enhanced,sarbicki2020family,terhal2000bell,guhne2009entanglement,ganguly2009witness,ganguly2013common,ganguly2014witness,goswami2019universal,aggarwal2024entanglement,bhattacharya2021generating,mallick2024genuine}. In the qubit-qubit and qubit-qutrit scenarios, the positive partial transposition (PPT) criterion serves as a necessary and sufficient condition for detecting entanglement \cite{horodecki2001separability}. However, in higher dimensions, the situation is more involved. Determining whether an arbitrary density matrix is separable has been proven to be NP-hard \cite{gurvits2004classical,gharibian2008strong,gharibian2015quantum}.\\

To gain a deeper insight into the quantum relationship between two particles, it is important not only to determine whether they are entangled but also to explore how many of their degrees of freedom contribute to the entanglement. This brings forth the concept of entanglement dimension, which can be quantified by the Schmidt number of a bipartite density matrix \cite{terhal2000schmidt}. The Schmidt number of a bipartite density matrix measures how many levels actively participate in generating the entanglement between the particles. In recent years, it has been shown that high dimensional entangled states i.e. quantum states with higher Schmidt numbers provide more advantages in several quantum information processing tasks such as channel discrimination \cite{bae2019more},  quantum
communication \cite{zhang2025quantum,cozzolino2019high}, quantum control \cite{kues2017chip}, universal quantum computation \cite{wang2020qudits},  experimental optics \cite{erhard2018twisted,erhard2020advances}. Further, in quantum key distribution, high dimensional entanglement can enhance key rates and improve error tolerance, thus enhancing security \cite{cerf2002security,sheridan2010security,cozzolino2019orbital,
bouchard2018experimental}. \\

Although states with higher Schmidt numbers offer several advantages, the most fundamental question is how to effectively detect the signature of Schmidt number of a given state. In \cite{terhal2000schmidt}, Terhal et. al. provide a criterion to detect the Schmidt number of a state by calculating the fidelity between the density matrix and the maximally entangled $d$-dimensional state. An alternative approach involves extending the well-known positive partial transpose (PPT) criterion \cite{horodecki2001separability,peres1996separability} and the computable cross-norm (CCNR) criterion \cite{chen2003matrix,rudolph2005further} to incorporate Schmidt numbers. Later, prescriptions based on covariance matrix\cite{liu2023characterizing,liu2024bounding} and Bloch decomposition \cite{klockl2015characterizing} have been developed to quantify Schmidt numbers. Recently, in \cite{tavakoli2024enhanced} authors proposed two simple criteria to detect the signature of the Schmidt number which are based on analyzing the correlations obtained from locally performing symmetric informationally complete (SIC) measurements and mutually unbiased bases (MUBs) quantum measurements. A witness-based method has also been developed to certify the Schmidt number of a state \cite{sanpera2001schmidt,shi2024families,zhang2024analyzing,wyderka2023construction,bavaresco2018measurements}. However, similar to the challenges associated with entanglement, characterizing the Schmidt number of an arbitrary bipartite density matrix still remains a challenging task in higher dimensional systems.\\

The evolution of quantum systems is represented by quantum channels. A quantum channel is a trace-preserving, completely positive map. There are specific scenarios in which the noise in such an environment can be sufficiently strong that the state loses its utility as a resource. Notable examples of such channels include non-locality breaking channel \cite{pal2015non}, negative conditional entropy breaking channel \cite{srinidhi2024quantum}, coherence breaking channel \cite{luo2022coherence}, teleportation resource breaking channel \cite{muhuri2023information}, incompatibility breaking channel \cite{heinosaari2015incompatibility}, steerability breaking channel \cite{ku2022quantifying}, magic breaking channel \cite{patra2024qubit} etc. Apart from these, there are certain classes of quantum channels that completely disentangle the subsystem they are acting on from the remainder of the system i.e. when these channels are applied to an arbitrary state, then the resulting state always becomes separable. These channels are called entanglement breaking channels \cite{horodecki2003entanglement}. A new family of quantum channels was
introduced in \cite{moravvcikova2010entanglement}, which act on a subsystem $B$ and completely destroy entanglement within subsystem $B$, but not necessarily destroy entanglement between the composite system $AB$. These channels are known as entanglement annihilating channels.\\

There is a class of channels that generalizes the notion of entanglement breaking channels, {\it viz.}, channels that reduce the Schmidt number of a bipartite composite state. These channels are known as partially entanglement breaking channels \cite{chruscinski2006partially}. Hence, partially entanglement breaking channels can be understood as resource breaking channels. Therefore, a fundamental and important question that arises from the resource theory perspective is how to identify the signature of a non-partially entanglement breaking channel \cite{mukherjee2025certifying}. In this work, we propose a method for the efficient detection of non-partially entanglement breaking channels. We then delve into the characterization of partially entanglement breaking channels, examining their structural properties and behavior. Furthermore, we introduce the concept of partially entanglement annihilating channels, offering a topological characterization and detailed exploration of their properties. We explore the qutrit depolarizing channel, determining the precise parameter regime in which it acts as a non-partially entanglement annihilating channel. Finally, we examine the relationships between entanglement breaking, partially entanglement breaking, and partially entanglement annihilating channels.\\

The structure of this paper is outlined as follows: In Sec.\ref{s2}, we present key preliminary concepts regarding entanglement breaking, entanglement annihilating, and partially entanglement breaking channels. In Sec.\ref{s3}, we provide a characterization of partially entanglement breaking channels, alongside a discussion of their fundamental properties. We address the detection of non-partially entanglement breaking channels by employing a suitable witness operator. Following this, in Sec.\ref{s5} we first introduce the notion of partially entanglement annihilating channels and then examine their topological characteristics and other properties. We then provide comparisons between entanglement breaking, partially entanglement breaking, and partially entanglement annihilating channels in Section~\ref{s6}. Finally, we conclude the paper in Sec. \ref{s7}, offering insights into some prospects of future research directions.

\section{Notations and preliminaries} \label{s2}
\begin{table}[ht] 
\centering

\begin{tabular}{| c| c| } 

\hline
 \textbf{Description}  &   \textbf{Notations}\\

 \hline
$d$ dimensional Complex Hilbert space & ${\mathbf{C}}^d$\\

\hline 
Identity matrix & $I$\\

\hline 
Identity map & $id_A$\\
\hline 
Choi Matrix & $\mathcal{C}$\\

 \hline
 Entanglement breaking channel  & $\Phi$\\
 
  \hline
  Set of all density matrices which has Schmidt number $\le$ r & $S_r$\\

  \hline 

  Entanglement annihilating channel & $\Phi_{A}$\\

  \hline 
partially entanglement breaking channel  &  $\mathcal{S}$\\

\hline 
partially entanglement annihilating channel  &  $\mathcal{E}$\\

\hline
Set of all entanglement breaking channels & $\mathbb{EB}$\\

\hline
Set of all entanglement annihilating channels & $\mathbb{EA}$\\

\hline
 Set of all r-partially entanglement breaking channels &  $r-\mathbb{PEBT}$\\

 \hline 
 Set of all r-partially entanglement annihilating channels &  $r-\mathbb{PEAT}$\\

  \hline 
\end{tabular}\\
\caption{The table depicts the different notations used in the work}   
\label{notationstable}
\end{table}


Table \ref{notationstable} presents the list of notations used throughout this work. Below, we discuss important prerequisites pertinent to our study.

 Let us consider a bipartite system, comprising subsystems $A$ and $B$ associated with the Hilbert space ${\mathcal{H}}_A \otimes {\mathcal{H}}_B$. 
 $\mathcal{B}({\mathcal{H}}_{AB})$ is the set of all bounded operators acting on Hilbert space ${\mathcal{H}}_A \otimes {\mathcal{H}}_B$.  The states associated with the Hilbert space are called density operators which are positive operators with unit trace. We denote the set of all density operators as $\mathcal{D} ({\mathcal{H}}_{AB})$.
 The states defined on the Hilbert space ${\mathcal{H}}_A  \otimes {\mathcal{H}}_B$ can be categorized into two subsets: separable and entangled. A separable state is a quantum state of a composite system ${\mathcal{H}}_A  \otimes {\mathcal{H}}_B$ that can be expressed as a tensor product of the states of its subsystems $A$ and $B$, i.e. 
\begin{equation}
    \rho_{AB}= \sum_{i} p_i {\rho_{A}}^i \otimes {\rho_{B}}^i
\end{equation}
where, $ p_i \ge 0$ and $\sum_{i} p_i =1$ and ${\rho_{A}}^i, {\rho_{B}}^i$ are the density matrices representing the states of subsystems $A$ and $B$. If a composite state can not be written in the above form, then the composite state is said to be entangled. 

Let, ${\mathcal{H}}_A = {\mathcal{H}}_B = {\mathbf{C}}^d$. The operators acting on a finite-dimensional Hilbert space can be expressed as matrices. Let, $\mathcal{M}_d$ and $\mathcal{M}_k$ denote the set of $d \times d$ and  $k \times k$ complex matrices. A linear map $\Phi: \mathcal{M}_d \rightarrow \mathcal{M}_d $ is said to be positive, if $\Phi(\rho) \ge 0$, for all $\rho \in \mathcal{M}_d$. A linear map $\Phi: \mathcal{M}_d \rightarrow \mathcal{M}_d$ is said to be $k$ positive if the induced map $id_A \otimes \Phi :\mathcal{M}_k \otimes \mathcal{M}_d \rightarrow \mathcal{M}_k \otimes \mathcal{M}_d$ is positive for some $k \in \mathbf{N}$. A linear map $\Phi: \mathcal{M}_d \rightarrow \mathcal{M}_d $ is said to be completely positive if $id_A \otimes \Phi : \mathcal{M}_k \otimes \mathcal{M}_d \rightarrow \mathcal{M}_k \otimes \mathcal{M}_d$ is positive for all $k \in \mathbf{N}$. To show that, a positive map $\Phi$ is completely positive, one makes use of the Choi-Jamiolkowski isomorphism \cite{jamiolkowski1974effective,choi1975positive}. The Choi matrix corresponding to the positive map $\Phi$ is $ {\mathcal{C}}_{\Phi} = (id_A \otimes \Phi) (\ket{{\phi}^+}\bra{{\phi}^+})$, where $\ket{{\phi}^+}=\frac{1}{\sqrt d} \sum_{i} \ket{ii}$ is the maximally entangled state in ${\mathbf{C}}^d\otimes {\mathbf{C}}^d$. Now, a positive map $\Phi$ is said to be completely positive iff the Choi matrix ${\mathcal{C}}_{\Phi}$ corresponding to the map is positive semidefinite. Furthermore, a linear map $\Phi$ is said to be trace-preserving if $\text{Tr} (\Phi (\rho)) = \text{Tr}(\rho)$ for all $\rho \in  \mathcal{M}_d$.

Quantum systems evolve through quantum channels, which are completely positive and trace-preserving maps (CPTP). These channels map any element from the set of all density operators into another element within the same set. Now, from Kraus-Choi representation theorem \cite{kraus1983states}, it is known that any CPTP map $\Phi: \mathcal{M}_d \rightarrow \mathcal{M}_d$ can be written as \begin{equation}
      \Phi (\rho) = \sum_{i} \mathcal{K}_{\alpha} \hspace{0.05cm}\rho \hspace{0.05cm}{{\mathcal{K}}_{\alpha}^{\dagger}}
  \end{equation}
  where, $\mathcal{K}_{\alpha} \in \mathcal{M}_d $ are the Kraus operators corresponding to the channel $\Phi$ satisfying $\sum_{\alpha} {{\mathcal{K}}^{\dagger}}_{\alpha}  \mathcal{K}_{\alpha} = I_d$.

\subsection{Entanglement breaking and annihilating channels}  
In the resource theory of entanglement, entangled states are resourceful states. There are certain classes of quantum channels that completely disentangle the subsystem they are acting on from the remainder of the system i.e. when these channels are applied to one part of an arbitrary state, then the resulting state always becomes separable. These channels are called entanglement breaking channels. Mathematically, we can say that, a channel $\Phi$ is entanglement breaking, 
if $(id_A \otimes \Phi) (\rho)$ is separable for all $\rho$ \cite{horodecki2003entanglement}. Later in \cite{horodecki2003entanglement}, authors proved that a quantum channel  $\Phi: \mathcal{M}_d \rightarrow \mathcal{M}_d $ is said to be entanglement breaking iff $(id_A \otimes \Phi) (\ket{{\phi}^{+}} \bra{{\phi}^{+}})$ is separable, where, $\ket{{\phi}^{+}}=\frac{1}{\sqrt d} \sum_{i} \ket{ii}$ is the maximally entangled state in ${\mathbf{C}}^d\otimes {\mathbf{C}}^d$  \hspace{0.1cm}\cite{horodecki2003entanglement}. It is known that \cite{horodecki2003entanglement}, an entanglement breaking channel can be written in terms of rank one Kraus operators. Moreover, an entanglement breaking channel can be interpreted as a measure and prepare scenario i.e. every entanglement breaking channel can be written as :
  \begin{equation}
  \Phi(\rho) = \sum_{i} \text{Tr}[ \rho F_i ] \hspace{0.05cm} \sigma_i
  \end{equation}
where $F_i$ are some positive operators-valued measurements (POVM) i.e. $F_i \ge 0$,  $\sum_i F_i =I_d$ and $\{\sigma_i\}$'s are some fixed states. 




    It is worth noting that an entanglement-breaking channel completely destroys entanglement across the bipartition $A|B$. In contrast, a distinct class of quantum channels, known as entanglement-annihilating channels, has been introduced more recently \cite{moravvcikova2010entanglement}. These channels completely destroy the entanglement within a given subsystem $B$, while not necessarily destroying the entanglement between subsystems $A$ and $B$. More precisely, let $\Phi_B$ be a quantum channel acting on subsystem $B$. Then $\Phi_B$ is said to be an entanglement-annihilating channel if, for every state $\rho_B \in \mathcal{D}(\mathcal{H}_B)$, the output state $\Phi_B(\rho_B)$ is separable \cite{moravvcikova2010entanglement,filippov2012local}.


\subsection{Partially entanglement breaking channels} Before discussing the partially entanglement breaking channels, let us first define the Schmidt rank of a pure state. 

A bipartite pure state can always be written in its Schmidt decomposition form as \cite{peres1997quantum}:
\begin{equation}
    \ket{\psi}= \sum_{i=1}^r \sqrt{{\lambda}_i} {\ket{i}}_A {\ket{i}}_B \nonumber
\end{equation}
where, ${\lambda}_i \ge 0$, $ \sum_i {\lambda}_i =1$, and $\ket{i_A} (\ket{i_B})$ forms an orthonormal basis in ${\mathcal{H}}_A ({\mathcal{H}}_B)$. Here, $r$ indicates the Schmidt rank of the pure state $ \ket{\psi}$. Later, Terhal and Horodecki \cite{terhal2000schmidt} extended this concept to mixed states by introducing the notion of Schmidt number (SN). A bipartite density matrix $\rho$ has Schmidt number $r$, if in every possible decomposition of $\rho$ into pure states, i.e. $\rho= \sum_{k} p_k {\ket{{\psi}_k}} {\bra{{\psi}_k}}$, with $p_k \ge 0$, at least one of the pure states ${\ket{{\psi}_k}}$ has Schmidt rank $r$. Moreover, there must exist at least one decomposition of $\rho$  in which all vectors  $\{\ket{{\psi}_k} \}$ have a Schmidt rank not greater than $r$. This can be expressed mathematically as:
\begin{equation}
    \text{SN} (\rho) = \min_{\rho = \sum_{k} p_{k}  \ket{{\psi}_k} \bra{{\psi}_k}}  \hspace{0.1cm}\{\max_{k} \hspace{0.1cm} \text{SR}(\ket{{\psi}_k)}\}
\end{equation}
where, the minimization is taken over all possible pure state decomposition of $\rho$ and SR $(\ket{{\psi}_k)}$ represents the Schmidt rank of the pure state $\ket{{\psi}_k}$. For a bipartite state $\rho \in  \mathcal{D}({{\mathcal{C}}^d \otimes  {\mathcal{C}}^d})$, the Schmidt number satisfies $1 \le$ \hspace{0.08cm}SN$(\rho) \le d$. If a bipartite quantum state $(\rho)$ is separable, then SN$(\rho) =1$. 

There are certain classes of quantum channels that reduce the Schmidt number of a bipartite composite state. These channels are known as partially entanglement breaking channels. Mathematically, A quantum channel $\mathcal{S}$ is said to be a r-partially entanglement breaking channel, iff SN$[(id_A \otimes \mathcal{S}) \rho] \le r$, for all $\rho \in  \mathcal{D}({{\mathbf{C}}^d \otimes  {\mathbf{C}}^d})$, where $r < d$ \cite{chruscinski2006partially,devendra2023mapping}. In \cite{chruscinski2006partially} Chruscinski et.al. proved that, a channel $\mathcal{S} \in$ $r-\mathbb{PEBT}$ iff SN$({\mathcal{C}}_{\mathcal{S}}) \le r$,  where $ {\mathcal{C}}_{\mathcal{S}} = (id_A \otimes \mathcal{S}) (\ket{{\phi}^{+}} \bra{{\phi}^{+}})$ with  $\ket{{\phi}^{+}}=\frac{1}{\sqrt d} \sum_{i} \ket{ii}$ is the maximally entangled state in ${\mathbf{C}}^d\otimes {\mathbf{C}}^d$.
We now proceed to characterize $r-\mathbb{PEBT}$ and reveal some of its important properties.
\section{Characterization and properties of $r-\mathbb{PEBT}$} \label{s3}
In this section, we present the characterization and properties of channels belonging to $r-\mathbb{PEBT}$. We begin with their characterization, where we establish structural features such as convexity, compactness, and detection criteria. We then proceed to investigate their properties, focusing on their behavior under different types of concatenation.

\subsection{Characterization of $r-\mathbb{PEBT}$}
\begin{lemma} \label{lemma1}
    $r-\mathbb{PEBT}$ is convex.
\end{lemma}

\textit{Proof:} To prove that the set $r-\mathbb{PEBT}$ is convex, we begin by assuming that both $\mathcal{S}_{1}$ and $\mathcal{S}_{2}$ are elements of $r-\mathbb{PEBT}$. Our goal is to show that for any $p \in [0,1]$, the convex combination $\mathcal{S}= p \mathcal{S}_{1} + (1-p) \mathcal{S}_{2} $, also belongs to $r-\mathbb{PEBT}$. \\

Since, $\mathcal{S}_{1} $,  $\mathcal{S}_{2} \in$ $r-\mathbb{PEBT}$, therefore 
\begin{equation}
    \text{SN} \Big( (id_A \otimes \mathcal{S}_{1}) \ket{{\phi}^+} \bra{{\phi}^+} \Big) \leq r \implies \text{SN} (\rho_1) \le r
\end{equation}
and 
\begin{equation}
     \text{SN} \Big( (id_A \otimes \mathcal{S}_{2}) \ket{{\phi}^+} \bra{{\phi}^+} \Big)  \le r \implies \text{SN} (\rho_2) \le r
\end{equation}
where, $ (id_A \otimes \mathcal{S}_{1}) \ket{{\phi}^+} \bra{{\phi}^+} = \rho_1$, $ (id_A \otimes \mathcal{S}_{2}) \ket{{\phi}^+} \bra{{\phi}^+} = \rho_2$ and  $\ket{{\phi}^{+}}=\frac{1}{\sqrt d} \sum_{i} \ket{ii}$.\\


 Now, 
\begin{align}
    &  \Big((id_A \otimes \mathcal{S}) \ket{{\phi}^+} \bra{{\phi}^+} \Big) \\ \nonumber
       =  & \Big( (id_A \otimes  (p \mathcal{S}_{1} + (1-p) \mathcal{S}_{2} ))\ket{{\phi}^+} \bra{{\phi}^+} \Big)\\ \nonumber
        = & p \hspace{0.15cm}  \rho_1 + (1-p) \hspace{0.15cm}  \rho_2   \nonumber
   \end{align}
 Since $\rho_1, \rho_2 \in S_r$ and $S_r$ forms a convex set \cite{terhal2000schmidt},  therefore, their convex combination $ ( p \hspace{0.15cm}  \rho_1 + (1-p) \hspace{0.15cm}  \rho_2)$ also belongs to $S_r$ which implies  SN$\Big((id_A \otimes \mathcal{S}) \ket{{\phi}^+} \bra{{\phi}^+} \Big) \leq r$. Therefore, $\mathcal{S} \in$  $r-\mathbb{PEBT}$. This completes the proof.\\

\begin{lemma} \label{lemma2}
    $r-\mathbb{PEBT}$ is compact.
\end{lemma} 

\textit{Proof:} First, let us prove that the set $r-\mathbb{PEBT}$ is closed, i.e. it contains all of its limit points. Let, $\mathcal{S}_{0}$ be an arbitrary limit point of $r-\mathbb{PEBT}$ (the set $r-\mathbb{PEBT}$ always has a limit point since we have proved earlier that the set is convex), therefore, if we consider an open ball $B_{\frac{1}{m}}(\mathcal{S}_{0})$ of radius $\frac{1}{m}$ centered on $\mathcal{S}_{0}$, then 
\begin{equation}
   \{ B_{\frac{1}{m}}(\mathcal{S}_{0}) - \mathcal{S}_{0} \} \cap \text{$r-\mathbb{PEBT}$} \neq \emptyset
\end{equation}
Since, each neighborhood of $\mathcal{S}_{0}$  contains infinitely many points of $r-\mathbb{PEBT}$, where $\emptyset$ indicates the null set.
Let us now construct a sequence $\{\mathcal{S}_m\}$ of distinct r-partially entanglement breaking channels such that   $\mathcal{S}_m \rightarrow \mathcal{S}_{0} $ as follows:
\begin{align}
    & \mathcal{S}_1 \in B_1(\mathcal{S}_{0}) \cap \text{$r-\mathbb{PEBT}$} , \hspace{0.15cm} \mathcal{S}_1 \neq \mathcal{S}_{0}\\ \nonumber
     &  \mathcal{S}_2 \in  B_{\frac{1}{2}}(\mathcal{S}_{0}) \cap \text{$r-\mathbb{PEBT}$} ,  \hspace{0.15cm} \mathcal{S}_2 \neq \mathcal{S}_{0},  \mathcal{S}_1 \\ \nonumber
      &   .... \in ....  \\ \nonumber
       &   .... \in ....  \\ \nonumber
         &  \mathcal{S}_m \in  B_{\frac{1}{m}}(\mathcal{S}_{0}) \cap \text{$r-\mathbb{PEBT}$} , \hspace{0.15cm}  \mathcal{S}_m \neq  \mathcal{S}_{0},...,\mathcal{S}_{m-1}
   \end{align}
From the above construction, it is evident that $\mathcal{S}_m \rightarrow \mathcal{S}_{0}$. Now, we take another sequence $\{ (id_A \otimes \mathcal{S}_m) (\rho) = \mu_m\}$ for arbitrary density operator $\rho$. Since $\mathcal{S}_m \in $ $r-\mathbb{PEBT}$, therefore, 
\begin{equation}
    \text{SN} [(id_A \otimes \mathcal{S}_m) (\rho)] = \text{SN} (\mu _m) \leq r
\end{equation}
We know that the set of states $(S_r)$ whose Schmidt number is less than equal to $r$ forms a closed set \cite{terhal2000schmidt}, therefore for $\mu _m \in S_r $, all limit points of $\mu _m$ also belongs to $ S_r$. Now, 
\begin{align}
   & \lim_{m\to\infty} (id_A \otimes \mathcal{S}_m) (\rho) \rightarrow (id_A \otimes \mathcal{S}_0) (\rho)  \\ \nonumber
   & \implies \lim_{m\to\infty} \mu _m = \mu _0   \hspace{1cm} [\text{where,} \hspace{0.2cm} (id_A \otimes \mathcal{S}_0) (\rho) = \mu_0]
\end{align}
Therefore, 
$\mu _0 \in  S_r$, which implies 
\begin{equation}
    \text{SN} [(id_A \otimes \mathcal{S}_0) (\rho)] \leq r
\end{equation}
i.e. $\mathcal{S}_0 \in $ $r-\mathbb{PEBT}$. Since we have considered $\mathcal{S}_0$ to be an arbitrary limit point of $r-\mathbb{PEBT}$, it follows that $r-\mathbb{PEBT}$ contains all of its limit points. Therefore, the set $r-\mathbb{PEBT}$ is closed.

Again, $r-\mathbb{PEBT}$ $\subseteq$ set of all CPTP maps. We know that any closed subset of a compact set is compact. Since the set of all CPTP maps forms a compact set, therefore the set $r-\mathbb{PEBT}$ is compact.

 \subsection{Detection of non-$r$-partially entanglement breaking channels} \label{s3b}
 In this section, we aim to identify quantum channels that are not $r$-partially entanglement breaking, as such channels play a crucial role in preserving high-dimensional entanglement, which is essential for quantum communication, channel discrimination, and several other quantum information processing tasks \cite{bae2019more,zhang2025quantum,cozzolino2019high,kues2017chip}. We begin by detecting $r$-partially entanglement breaking channels using witness operators and subsequently provide a sufficient condition for a channel to belong to $r\text{-}\mathbb{PEBT}$.

\subsubsection{Detection using witness operators :}
Here, our goal is to identify channels that are not $r$-partially entanglement breaking ($r\text{-}\mathbb{PEBT}$), i.e., channels capable of preserving Schmidt number strictly greater than $r$. To this end, we employ witness-based framework, where witnesses are constructed using a general approach based on distance-type functionals to detect channels outside the set $r\text{-}\mathbb{PEBT}$. Since the set $r\text{-}\mathbb{PEBT}$ is convex and compact (as established in Lemma \eqref{lemma1} and Lemma \eqref{lemma2}), this structure guarantees that such channels can be effectively identified  \cite{holmes2012geometric}. In particular, consider the following scalar-valued functional:
\begin{equation}
W(\Psi) = \max \left\{ 0, \inf_{\mathcal{S} \in r\text{-PEBT}} \|\Psi - \mathcal{S}\|_\diamond \right\},
\end{equation}
where $\|\cdot\|_\diamond$ denotes the diamond norm on quantum channels. By definition, if $\mathcal{S} \in r\text{-}\mathbb{PEBT}$, then $W(\mathcal{S}) = 0$, whereas $W(\Psi) > 0$ implies that $\Psi \notin r\text{-}\mathbb{PEBT}$. By the compactness of the set $r\text{-}\mathbb{PEBT}$, there exists at least one channel $\mathcal{S}^\ast \in r\text{-}\mathbb{PEBT}$ that minimizes the diamond-norm distance to $\Psi$. Consequently, the functional $W(\Psi)$ serves as a witness capable of identifying channels that are not $r$-partially entanglement breaking. It is worth noting, however, that identifying the closest $r$-partially entanglement breaking channel is computationally demanding. Nevertheless, this formulation offers a clear and broadly applicable framework for detecting channels that preserve higher Schmidt number.


 Below, we present an alternative approach for detecting channels that are not r-partially entanglement breaking. Specifically, we employ a witness operator $W$ acting on the Choi state of the channel. Such a witness is constructed to satisfy $\text{Tr}(\mathcal{W} \hspace{0.1cm} (id_A \otimes {\mathcal{S}}) \ket{{\phi}^+} \bra{{\phi}^+}) \ge 0$ for all ${\mathcal{S}} \in$ $r-\mathbb{PEBT}$ and $\text{Tr}(\mathcal{W} \hspace{0.1cm} (id_A \otimes {\mathcal{N}}) \ket{{\phi}^+} \bra{{\phi}^+}) < 0$ for atleast one ${\mathcal{N}} \notin$ $r-\mathbb{PEBT}$. This approach provides a practically motivated alternative to distance-based methods, particularly in scenarios where computing the diamond norm is intractable. To illustrate the effectiveness of this approach, we present two examples involving the qutrit depolarizing and dephasing channels. In each case, we explicitly identify the parameter regimes for which these channels fail to be $2$-partially entanglement breaking, thereby demonstrating how the witness-based method accurately detects the range over which the channels lie outside $2\text{-}\mathbb{PEBT}$. \\

\textbf{Depolarizing channel:} Let us first consider the depolarizing channel $  {\mathcal{S}}_d : \mathcal{M}_d \rightarrow \mathcal{M}_d $ whose action is given by   \cite{nielsen2001quantum,chruscinski2006partially},
\begin{equation} \label{depolarizing}
    {\mathcal{S}}_d(\rho) = p \rho + \frac{1-p}{d} Tr(\rho) I_d
\end{equation}
where, $p\in [0,1]$. It is known that \cite{terhal2000schmidt}, SN$( {\mathcal{S}}_d) \le r$ iff 
\begin{equation} \label{SNphi}
  0 \le  p \le  \frac{rd-1}{d^2-1}
\end{equation} 

Now, an optimal witness to detect states with SN $>$ r is \cite{sanpera2001schmidt},
\begin{equation} \label{witness}
    \mathcal{W} = I_{d^2} - \frac{d}{r} \hspace{0.1cm}\mathcal{P}
\end{equation}
where, $\mathcal{P}$ is a projector onto a maximally entangled state $\ket{{\phi_d}^+} = \sum_{i=1}^d \frac{1}{\sqrt{d}} \ket{ii}$.


Let us take $d=3$. Then Eq.\eqref{depolarizing} becomes
\begin{equation}
   {\mathcal{S}}_3 (\rho) = p \rho + \frac{1-p}{3} Tr(\rho) I_3
\end{equation}
From Eq.\eqref{SNphi}, it follows that $ {\mathcal{S}}_3 \in$ $2-\mathbb{PEBT}$ iff $ 0 \le p \le  \frac{5}{8}$ and SN $({\mathcal{S}}_3) =3$ iff $  \frac{5}{8} < p \le 1 $. Using the optimal witness defined in Eq. \eqref{witness}, we get 
\begin{equation}
    \text{Tr}(\mathcal{W} \hspace{0.1cm} (id_A \otimes {\mathcal{S}}_3) \ket{{\phi_3}^+} \bra{{\phi_3}^+}) \ge 0 \hspace{0.2cm} \text{for} \hspace{0.2cm} 0 \le p \le  \frac{5}{8}
\end{equation}
and 
\begin{equation}
    \text{Tr}(\mathcal{W} \hspace{0.1cm} (id_A \otimes {\mathcal{S}}_3) \ket{{\phi_3}^+} \bra{{\phi_3}^+}) < 0 \hspace{0.2cm} \text{for} \hspace{0.2cm} \frac{5}{8} < p \le 1  
\end{equation}
Hence, $\mathcal{W}$ can detect the full range of the parameter $p$ for which $ {\mathcal{S}}_3 \notin 2-\mathbb{PEBT}$ i.e. SN$({\mathcal{S}}_3) =3$.\\

\textbf{Dephasing Channel:} Here we consider the dephasing channel $  {\mathcal{D}}_d : \mathcal{M}_d \rightarrow \mathcal{M}_d $ whose action is given as \cite{nielsen2001quantum,tavakoli2024enhanced},
\begin{equation} \label{dephasing}
    {\mathcal{D}}_d(\rho) = v \rho + (1-v) \sum_{i=0}^{d-1} \ket{i}\bra{i}\rho\ket{i}\bra{i}
\end{equation}
where, $v\in [0,1]$. It is known that \cite{tavakoli2024enhanced,bavaresco2018measurements}, SN$( {\mathcal{D}}_d) \le r$ iff
\begin{equation} \label{SND}
   0 \le v \le \frac{r-1}{d-1}
\end{equation}
We focus on the qutrit dephasing channel by setting $d=3$ in Eq.\eqref{dephasing}. From Eq.\eqref{SND}, it follows that $ {\mathcal{D}}_3 \in 2-\mathbb{PEBT}$ iff $ 0 \le v \le  \frac{1}{2}$ and SN $( {\mathcal{D}}_3) =3$, i.e. $ {\mathcal{D}}_3 \notin 2-\mathbb{PEBT}$ iff $  \frac{1}{2} < v \le 1 $. Using the optimal witness defined in Eq. \eqref{witness}, we obtain
\begin{equation}
    \text{Tr}(\mathcal{W} \hspace{0.1cm} (id_A \otimes {\mathcal{D}}_3) \ket{{\phi_3}^+} \bra{{\phi_3}^+}) \ge 0 \hspace{0.2cm} \text{for} \hspace{0.2cm} 0 \le v \le  \frac{1}{2}
\end{equation}
and 
\begin{equation}
    \text{Tr}(\mathcal{W} \hspace{0.1cm} (id_A \otimes {\mathcal{D}}_3) \ket{{\phi_3}^+} \bra{{\phi_3}^+}) < 0 \hspace{0.2cm} \text{for} \hspace{0.2cm} \frac{1}{2} < v \le 1  
\end{equation}
Hence, $\mathcal{W}$ can detect the full range of the parameter $v$ for which the channel $ {\mathcal{D}}_3 \notin 2-\mathbb{PEBT}$  i.e. SN$({\mathcal{D}}_3) =3$.\\

\subsubsection{Sufficient condition to detect non-$r$-partially entanglement breaking channels :} Here, we establish a sufficient criterion for a quantum channel to belong to $r\text{-}\mathbb{PEBT}$.\\

\begin{theorem}
    If for every linear $r$-positive but not $(r+1)$-positive map $(\Lambda)$, a quantum channel $\mathcal{S}$ satisfies
\begin{equation}
    \operatorname{Tr}\!\left[(\mathrm{id}_A \otimes \Lambda)\big(\mathcal{C}_{\mathcal{S}}\big)\right]^2 \le \frac{1}{d^2 - 1},
\end{equation}
where $\mathcal{C}_{\mathcal{S}}$ denotes the Choi operator of $\mathcal{S}$, then the channel $\mathcal{S}$ belongs to $r-\mathbb{PEBT}$.
\end{theorem}

\textit{Proof:} To prove that $\mathcal{S}$ belongs to $r-\mathbb{PEBT}$, we have to prove that
\begin{equation}
    (\mathrm{id}_A \otimes \Lambda)\big(\mathcal{C}_{\mathcal{S}}\big) \ge 0
\end{equation}

for all $r$-positive but not $(r+1)$-positive map $\Lambda$ . We proceed by contradiction.

Suppose there exists a $r$-positive but not $(r+1)$-positive map $\Lambda$ such that
\begin{equation} \label{initialassumption}
    (\mathrm{id}_A \otimes \Lambda)\big(\mathcal{C}_{\mathcal{S}}\big) \ngeq 0
\end{equation}
i.e., the operator has at least one negative eigenvalue. Let $\{\zeta_i\}_{i=1}^{d^2}$ denote the eigenvalues of this operator, and assume without loss of generality that $\zeta_1 < 0$. Now,
\begin{align*}
     \operatorname{Tr}\!\left[(\mathrm{id}_A \otimes \Lambda)\big(\mathcal{C}_{\mathcal{S}}\big)\right] &= \sum_{i=1}^{d^2} \zeta_i \\
    & = \zeta_1 + \sum_{i=2}^{d^2} \zeta_i \\
    & <  \sum_{i=2}^{d^2} \zeta_i  \hspace{0.4cm}[\text{Since,} \hspace{0.1cm} \zeta_1 < 0]\\
    & \overset{a} \le \left(\sum_{i=2}^{d^2} \zeta_i^2\right)^{1/2}
\left(\sum_{i=2}^{d^2} 1^2\right)^{1/2} \\
    &  = \sqrt{d^2-1} \left( \sum_{i=2}^{d^2} \zeta_i^2 \right)^{1/2} \\ 
    & < \sqrt{d^2-1} \left( \sum_{i=1}^{d^2} \zeta_i^2 \right)^{1/2} = \sqrt{d^2-1} \, \left[\mathrm{Tr}[(\mathrm{id}_A \otimes \Lambda)\big(\mathcal{C}_{\mathcal{S}}\big)]^2\right]^{\frac{1}{2}}
\end{align*}
where ($a$) follows from the Cauchy–Schwarz inequality \cite{bhatia2013matrix}. Therefore,
 \begin{equation}
      \operatorname{Tr}\!\left[(\mathrm{id}_A \otimes \Lambda)\big(\mathcal{C}_{\mathcal{S}}\big)\right]  < \sqrt{d^2-1} \, \left[\mathrm{Tr}[(\mathrm{id}_A \otimes \Lambda)\big(\mathcal{C}_{\mathcal{S}}\big)]^2\right]^{\frac{1}{2}}
 \end{equation}
Since, $ \operatorname{Tr}\!\left[(\mathrm{id}_A \otimes \Lambda)\big(\mathcal{C}_{\mathcal{S}}\big)\right]  =1$, hence we obtain 
 \begin{equation}
     1  < \sqrt{d^2-1} \, \left[\mathrm{Tr}[(\mathrm{id}_A \otimes \Lambda)\big(\mathcal{C}_{\mathcal{S}}\big)]^2\right]^{\frac{1}{2}}
 \end{equation}
Squaring both sides, we obtain 
\begin{equation}
    \left[\mathrm{Tr}[(\mathrm{id}_A \otimes \Lambda)\big(\mathcal{C}_{\mathcal{S}}\big)]^2\right] > \frac{1}{d^2-1}
\end{equation}

Thus, our initial assumption [Eq.\eqref{initialassumption}] must be false, and it follows that for all $r$-positive but not $(r+1)$-positive map $\Lambda$ such that
\begin{equation} \label{initialassumption}
    (\mathrm{id}_A \otimes \Lambda)\big(\mathcal{C}_{\mathcal{S}}\big) \ge 0
\end{equation}
Therefore, the channel $\mathcal{S}$ belongs to $r-\mathbb{PEBT}$.\\

\subsection{Properties of $r-\mathbb{PEBT}$}
   In this subsection, we discuss several properties of channels belonging to $r-\mathbb{PEBT}$, with particular emphasis on their behavior under different types of concatenation. \\

\textbf{Series Concatenation:} Let $\mathcal{S}_1$ and $\mathcal{S}_2$ be two r-partially entanglement breaking channels. We denote the series concatenation of these two quantum channels by $\mathcal{S}_1 \circ \mathcal{S}_2$ and represent the action on input state $\rho$ as $\mathcal{S}_1 \circ \mathcal{S}_2 = \mathcal{S}_1(\mathcal{S}_2 (\rho))$.\\

\begin{theorem}
    Let, $\mathcal{S}_1 , \mathcal{S}_2 \in$ $r-\mathbb{PEBT}$, then $\mathcal{S}_1 \circ \mathcal{S}_2 \in$ $r-\mathbb{PEBT}$.
\end{theorem}  

\textit{Proof:} Since, $\mathcal{S}_1 \in $ $r-\mathbb{PEBT}$, therefore \begin{equation}
    \text{SN} \Big( (id_A \otimes \mathcal{S}_{1}) \ket{{\phi}^+} \bra{{\phi}^+} \Big) \leq r
\end{equation}
Again,  $\mathcal{S}_2 \in $ $r-\mathbb{PEBT}$, therefore \begin{equation}
    \text{SN} \Big( (id_A \otimes \mathcal{S}_{2}) \ket{{\phi}^+} \bra{{\phi}^+} \Big) \leq r
\end{equation}
Now, consider the composition $\mathcal{S}_{1} \circ \mathcal{S}_2$: \begin{equation}
     \Big( (id_A \otimes (\mathcal{S}_{1} \circ \mathcal{S}_2)) \ket{{\phi}^+} \bra{{\phi}^+} \Big) 
    =   \Big( (id_A \otimes \mathcal{S}_{1}) ( id_A \otimes \mathcal{S}_2)) \ket{{\phi}^+} \bra{{\phi}^+} \Big)
   =    \Big( (id_A \otimes \mathcal{S}_{1}) \hspace{0.2cm} \sigma  \Big)  
    \end{equation}
 \text{where}, $\sigma = (id_A \otimes \mathcal{S}_{2})\ket{{\phi}^+} \bra{{\phi}^+}$. Since, $\mathcal{S}_{2} \in$ $r-\mathbb{PEBT}$, we have SN$(\sigma )\le r$. Thus,
   \begin{equation}
       \Big( (id_A \otimes (\mathcal{S}_{1} \circ \mathcal{S}_2)) \ket{{\phi}^+} \bra{{\phi}^+} \Big) = \rho \hspace{.7cm} [\text{where}, \rho = (id_A \otimes \mathcal{S}_{1}) \sigma]
   \end{equation}
Since, $\mathcal{S}_1 \in$ $r-\mathbb{PEBT}$, it follows that SN$(\rho) \le r$. Hence, SN$\Big( (id_A \otimes (\mathcal{S}_{1} \circ \mathcal{S}_2)) \ket{{\phi}^+} \bra{{\phi}^+} \Big) \le r$ which implies that $\mathcal{S}_1 \circ \mathcal{S}_2 \in$ $r-\mathbb{PEBT}$.\\\\
\textbf{Parallel Concatenation:} Entanglement breaking channels are known to be closed under the tensor product. In other words, if $\Phi_1: \mathcal{M}_d \rightarrow \mathcal{M}_d$ and $\Phi_2: \mathcal{M}_d \rightarrow \mathcal{M}_d$ are entanglement breaking channels, then their tensor product is also an entanglement breaking channel. Here, we aim to explore whether a similar property extends to partially entanglement breaking channels, specifically, whether the tensor product of two such channels yields a partially entanglement breaking channel. \\

\begin{observation} \label{observation1}
     Let, $\mathcal{S}_1 , \mathcal{S}_2 \in$ $r-\mathbb{PEBT}$, then $\mathcal{S}_1 \otimes \mathcal{S}_2$ may not belong to $r-\mathbb{PEBT}$.
\end{observation} 

\textit{Proof:} Since \(\mathcal{S}_1,\mathcal{S}_2 \in r\text{-}\mathbb{PEBT}\), thereofore there exist atleast one Kraus decompositions with Kraus operators\(\{\mathcal{K}_{\alpha}\}\) and \(\{\mathcal{R}_{\beta}\}\) corresponding to \(\mathcal{S}_1\) and \(\mathcal{S}_2\), respectively, such that \cite{chruscinski2006partially} 
\[
\operatorname{rank}(\mathcal{K}_{\alpha}) \leq r, 
\qquad 
\operatorname{rank}(\mathcal{R}_{\beta}) \leq r,
\]
for all \(\alpha,\beta\), with
\[
\sum_{\alpha} \mathcal{K}_{\alpha}^{\dagger}\mathcal{K}_{\alpha}=I_d,
\qquad
\sum_{\beta} \mathcal{R}_{\beta}^{\dagger}\mathcal{R}_{\beta}=I_d.
\]

Now, the tensor product channel \(\mathcal{S}_1 \otimes \mathcal{S}_2\) admits the Kraus representation 
\(\{\mathcal{K}_{\alpha}\otimes \mathcal{R}_{\beta}\}\). Using the identity
\begin{equation}
\operatorname{rank}(\mathcal{K}_{\alpha}\otimes \mathcal{R}_{\beta})
=
\operatorname{rank}(\mathcal{K}_{\alpha})\,
\operatorname{rank}(\mathcal{R}_{\beta}),
\end{equation}
we obtain
\begin{equation}
\operatorname{rank}(\mathcal{K}_{\alpha}\otimes \mathcal{R}_{\beta})
\leq r^2.
\end{equation}
Hence, the ranks of the Kraus operators associated with 
\(\mathcal{S}_1 \otimes \mathcal{S}_2\) may exceed \(r\). Therefore, in general, $\mathcal{S}_1\otimes \mathcal{S}_2 \notin r\text{-}\mathbb{PEBT}$. However, since every Kraus operator of the tensor product channel has rank at most \(r^2\), it follows that $\mathcal{S}_1\otimes \mathcal{S}_2 \in r^2\text{-}\mathbb{PEBT}.$

\begin{remark}
From Observation \eqref{observation1}, it follows that, unlike the class of entanglement breaking channels (\(r=1\)), the set of \(r\)-partially entanglement breaking channels is, in general, not closed under tensor products for \(r>1\). We further note that the above observation relies on the existence of suitable Kraus decompositions, whose explicit construction is generally mathematically intricate, making concrete counterexamples to Observation \eqref{observation1} particularly challenging.
\end{remark}

   \begin{proposition}
       If $\mathcal{S} \in$ $r-\mathbb{PEBT}$ and ${\mathcal{F}}_B$ be any quantum channel acting on subsystem B. Then, $\mathcal{S} \circ {\mathcal{F}}_B $ and ${\mathcal{F}}_B \circ \mathcal{S} $  both belongs to $r-\mathbb{PEBT}$.
   \end{proposition} 
   \textit{Proof:} Since, $\mathcal{S} \in $ $r-\mathbb{PEBT}$, we have \begin{equation}
    \text{SN} \Big( (id_A \otimes \mathcal{S}) (\rho) \Big) \leq r \hspace{0.2cm} \text{for all} \hspace{0.2cm} \rho \in \mathcal{D}{(\mathcal{C}}^d \otimes {\mathcal{C}}^d)
\end{equation}
   Now, consider the composite map $\mathcal{S} \circ {\mathcal{F}}_B $. We can express its action on $\ket{{\phi}^{+}}=\frac{1}{\sqrt d} \sum_{i} \ket{ii}$ as:
   \begin{equation}
     \Big( (id_A \otimes (\mathcal{S} \circ {\mathcal{F}}_B)) \ket{{\phi}^+} \bra{{\phi}^+} \Big)  =   \Big( (id_A \otimes \mathcal{S}) \circ ( id_A \otimes {\mathcal{F}}_B)) \ket{{\phi}^+} \bra{{\phi}^+} \Big) =   \Big( (id_A \otimes \mathcal{S}) \hspace{0.1cm} \sigma  \Big)   
    \end{equation}
where, $\sigma = (id_A \otimes {\mathcal{F}}_B)\ket{{\phi}^+} \bra{{\phi}^+}$. 
Since  $\mathcal{S} \in$ $r-\mathbb{PEBT}$, it follows that SN$ \Big( (id_A \otimes \mathcal{S}) \hspace{0.1cm} \sigma  \Big) \le r$. Thus, we conclude that $\mathcal{S} \circ {\mathcal{F}}_B \in$ $r-\mathbb{PEBT}$.\\

Next, we want to prove that ${\mathcal{F}}_B \circ \mathcal{S} \in$ $r-\mathbb{PEBT}$. The action of the channel ${\mathcal{F}}_B \circ \mathcal{S}$ can be written as 
 \begin{equation}
     \Big( (id_A \otimes ({\mathcal{F}}_B \circ \mathcal{S})) \ket{{\phi}^+} \bra{{\phi}^+} \Big) 
    =  \Big( (id_A \otimes {\mathcal{F}}_B) \circ ( id_A \otimes \mathcal{S})) \ket{{\phi}^+} \bra{{\phi}^+} \Big)
   =   \Big( (id_A \otimes {\mathcal{F}}_B) \hspace{0.2cm} \sigma  \Big)     
\end{equation}
where, $\sigma = (id_A \otimes \mathcal{S})\ket{{\phi}^+} \bra{{\phi}^+}$. Since,  $\mathcal{S} \in$ $r-\mathbb{PEBT}$, therefore SN$(\sigma) \le r$. Again, SN$( (id_A \otimes {\mathcal{F}}_B)  \sigma ) \le r$, because, the Schmidt number of a state can not increase under local operations \cite{terhal2000schmidt}.
Therefore, ${\mathcal{F}}_B \circ \mathcal{S} \in$ $r-\mathbb{PEBT}$.

\begin{proposition}
    $\mathcal{S}$ $\in$ $r-\mathbb{PEBT}$ iff adjoint of $\mathcal{S}$ $\in$ $r-\mathbb{PEBT}$ \cite{johnston2008partially}.
\end{proposition} 

\textit{Proof:} From  Kraus-Choi representation theorem \cite{kraus1983states}, we can represent $\mathcal{S} : \mathcal{M}_d \rightarrow \mathcal{M}_d $ as:
\begin{equation}
   \mathcal{S} (\rho) =  \sum_{\alpha} \mathcal{K}_{\alpha} \hspace{0.05cm}\rho \hspace{0.05cm}{{\mathcal{K}}_{\alpha}^{\dagger}}
\end{equation}
where, $\mathcal{K}_{\alpha} \in \mathcal{M}_d $ are the Kraus operators corresponding to the channel $\mathcal{S}$ satisfying $\sum_{\alpha} {{\mathcal{K}}^{\dagger}}_{\alpha}  \mathcal{K}_{\alpha} = I_d$. Since, $\mathcal{S}$ has Kraus $\{\mathcal{K} _{\alpha}\}$, therefore, adjoint of $\mathcal{S}$ i.e. ${\mathcal{S}}^{\dagger}$ has Kraus $\{{{\mathcal{K}}_{\alpha}^{\dagger}} \}$. It is known that \cite{chruscinski2006partially}, $\mathcal{S} \in$ $r-\mathbb{PEBT}$ if and only if there exists a Kraus decomposition of $\mathcal{S}$ such that all Kraus operators $\mathcal{K}_{\alpha}$ satisfy rank$(\mathcal{K} _{\alpha}) \le $ r, which implies rank$({{\mathcal{K}}_{\alpha}^{\dagger}}) \le $ r. Therefore, ${\mathcal{S}}^{\dagger} \in$ $r-\mathbb{PEBT}$.

    \section{Definition, characterization, and properties of $r$-partially entanglement annihilating channels} \label{s5}
    In this section, we will first define $r$-partially entanglement annihilating channels and then further examine their characteristics and properties. We know that r-partially entanglement breaking channels are those that decrease Schmidt number across the split $A|B$, whereas $r$-partially entanglement annihilating channels are those channels that reduce Schmidt number within subsystem B. The action of $r$-partially entanglement annihilating channels, consisting of two particles within system B, can be classified into two types: local and non-local. In the following subsection, we will first delineate the concepts of local and non-local $r$-partially entanglement annihilating channels. We will then provide a characterization of non-local partially entanglement annihilating channels. 

\subsection{Definitions of local and non-local $r$-partially entanglement annihilating channels}
A channel $\mathcal{E}$ is said to be local if it can be expressed as a tensor product of individual channels, i.e. $\mathcal{E} = {\mathcal{E}}_1 \otimes {\mathcal{E}}_2$. If a quantum channel can not be written in the above form, then it is said to be a non-local channel. \\

\textbf{Non-local $r$-partially entanglement annihilating channels :} Let, $\mathcal{E}$ be a quantum channel acting on subsystem B. We call $\mathcal{E}$ to be a non-local $r$-partially entanglement annihilating channel if $\mathcal{E}$ can not be written as ${\mathcal{E}}_1 \otimes {\mathcal{E}}_2$ and SN$({\mathcal{E}(\rho_{B})}) \le r$, for all $\rho_{B} \in \mathcal{D(H_B)}$, where, $r < d$ and $B$ is a bipartite system $\in \mathcal{D}({{\mathbf{C}}^d \otimes  {\mathbf{C}}^d})$.\\

\textbf{Local $r$-partially entanglement annihilating channels :} A single particle channel $(\mathcal{E})$ is said to be $2$-locally $r$-partially entanglement annihilating channel, if one can write $\mathcal{E}$ as ${\mathcal{E}}_1 \otimes {\mathcal{E}}_2$ and SN${[(\mathcal{E}}\otimes \mathcal{E}) ({\rho}_B)] \le r$, for all $\rho_{B} \in \mathcal{D(H_B)}$, where, $r < d$ and $B$ is a bipartite system $\in \mathcal{D}({{\mathbf{C}}^d \otimes  {\mathbf{C}}^d})$ \\

In this paper, we adopt the terminology non-local $r-\mathbb{PEAT}$ and $2$-locally $r-\mathbb{PEAT}$ to represent set of all non-local $r$-partially entanglement annihilating channels and set of all $2$-locally $r$-partially entanglement annihilating channels respectively.


\subsection{Characterization and properties of non-local $r-\mathbb{PEAT}$}
\begin{lemma} \label{lemma3}
    Non-local $r-\mathbb{PEAT}$ is convex.
\end{lemma} 

\begin{lemma} \label{lemma4}
    Non-local $r-\mathbb{PEAT}$ is compact.
\end{lemma}

 Proofs of Lemma \eqref{lemma3} and Lemma \eqref{lemma4} are provided in Appendix \eqref{appendixA}.\\

\textbf{Series Concatenation:} Let $\mathcal{E}_1$ and $\mathcal{E}_2$ be two non-local $r-\mathbb{PEAT}$. Next, we will prove that the series concatenation of two non-local $r$-partially entanglement annihilating channels is also a non-local $r$-partially entanglement annihilating channel.

\begin{theorem}
    Let, $\mathcal{E}_1 , \mathcal{E}_2 \in$ non-local $r-\mathbb{PEAT}$, then $\mathcal{E}_1 \circ \mathcal{E}_2 \in$ non-local $r-\mathbb{PEAT}$.
\end{theorem} 

\textit{Proof:} Since, ${\mathcal{E}}_1$, ${\mathcal{E}}_2 \in$ non-local $r-\mathbb{PEAT}$. Therefore, 
\begin{equation}
    \text{SN} ({\mathcal{E}}_1 ({\rho}_B)) \le r \hspace{0.5cm} \text{and} \hspace{0.5cm} \text{SN} ({\mathcal{E}}_2 ({\rho}_B)) \le r
\end{equation} for all ${\rho}_B \in \mathcal{D}({\mathcal{H}_B})$. Our goal is to show that the series concatenation $\mathcal{E}_1 \circ \mathcal{E}_2$, also belongs to non-local $r-\mathbb{PEAT}$.
    
Now, the action of $\mathcal{E}_1 \circ \mathcal{E}_2$ on arbitrary density operator ${\sigma}_B \in \mathcal{D}({\mathcal{H}_B})$ can be written as: 
\begin{equation}
  \Big(  (\mathcal{E}_{1} \circ \mathcal{E}_2) ({\sigma}_B)  \Big) 
    =  \Big( \mathcal{E}_{1}  (\mathcal{E}_2 ({\sigma}_B)) \Big) 
   =  \Big( \mathcal{E}_{1} ({\sigma}^2_B)  \Big)   =  {\sigma}^1_B   
   \end{equation}
where, ${\sigma}^2_B =   \mathcal{E}_2 ({\rho}_B)$ and  ${\sigma}^1_B = \mathcal{E}_{1} ({\sigma}^2_B)$. Since, $\mathcal{E}_{1} \in$ non-local $r-\mathbb{PEAT}$, therefore SN$({\sigma}^1_B) \le r$. Hence, $\mathcal{S}_1 \circ \mathcal{S}_2 \in$ non-local $r-\mathbb{PEAT}$.\\

\begin{proposition}
    If $\mathcal{E} \in$ non-local $r-\mathbb{PEAT}$ and ${\mathcal{F}}_B$ be any quantum channel acting on subsystem B. Then, $\mathcal{E} \circ {\mathcal{F}}_B $ belongs to non-local $r-\mathbb{PEAT}$ but ${\mathcal{F}}_B \circ \mathcal{E} $  may not belong to non-local $r-\mathbb{PEAT}$.
\end{proposition} 
   \textit{Proof:} Since,  $\mathcal{E} \in$ non-local $r-\mathbb{PEAT}$, we have \begin{equation}
   \text{SN} ({\mathcal{E}} ({\sigma}_B)) \le r \hspace{0.2cm} \text{for all} \hspace{0.2cm} {\sigma}_B \in \mathcal{D}({\mathcal{H}_B})
\end{equation}
   Now, consider the composite map $\mathcal{E} \circ {\mathcal{F}}_B $. We can express its action on an arbitrary state ${\rho}_B \in \mathcal{D}({\mathcal{H}_B})$  as:
   \begin{equation}
     \Big( (\mathcal{E} \circ {\mathcal{F}}_B) \hspace{0.1cm} {\rho}_B \Big) \\
    =    \Big( \mathcal{E}  ({\mathcal{F}}_B \hspace{0.1cm} ({\rho}_B ))\Big)\\ 
   =   \Big( \mathcal{E}  ({\sigma}_B) \Big)     \\ 
\end{equation}
where, ${\sigma}_B = {\mathcal{F}}_B \hspace{0.1cm} ({\rho}_B)$. Since  $\mathcal{E} \in$ non-local $r-\mathbb{PEAT}$, it follows that SN$({\mathcal{E}} ({\sigma}_B)) \le r$ for arbitrary ${\sigma}_B \in \mathcal{D}({\mathcal{H}_B})$. Thus, we conclude that $\mathcal{E} \circ {\mathcal{F}}_B \in$ non-local $r-\mathbb{PEAT}$.\\

Next, we want to check whether the combination ${\mathcal{F}}_B \circ \mathcal{E} $ also belongs to non-local $r-\mathbb{PEAT}$. \\
Now,
 \begin{equation}
     \Big( ({\mathcal{F}}_B \circ \mathcal{E}) \hspace{0.1cm} {\sigma}_B \Big) \\
    =   \Big(  {\mathcal{F}}_B (\mathcal{E} \hspace{0.1cm} ({\sigma}_B))\Big)\\ 
   =  \Big(  {\mathcal{F}}_B ({\rho}_B) \Big)    \\ 
   \end{equation}
    where, $\rho_B = \mathcal{E} \hspace{0.1cm} ({\sigma}_B)$. Since, ${\mathcal{F}}_B$ is an arbitrary channel acting on subsystem B only, therefore, Schmidt number of $\rho_B$ can increase under the action of ${\mathcal{F}}_B$,  For example, if ${\mathcal{F}}_B$ is a non-local unitary operation, then the Schmidt number of $\rho_B$ could increase. Hence, SN $\Big(  {\mathcal{F}}_B ({\rho}_B) \Big) $ may not be necessarily less than or equal to r i.e. ${\mathcal{F}}_B \circ \mathcal{E} $  may not belong to non-local $r-\mathbb{PEAT}$.\\
    
 For example, let $\mathcal{E}$ be a non-local $2$-partially entanglement breaking channel such that the resulting state has Schmidt number at most $2$. Consider the resulting state after the action of $\mathcal{E}$ as
\begin{equation}
\rho_B^2
= p \ket{\phi_2^+}\bra{\phi_2^+} + \frac{1-p}{9} I_9,
\end{equation}
where
\begin{equation}
\ket{\phi_2^+} = \frac{1}{\sqrt{2}}(\ket{00} + \ket{22}) \in \mathbb{C}^3 \otimes \mathbb{C}^3.
\end{equation}
Since the set of states with Schmidt number at most $2$ is convex \cite{terhal2000schmidt}, it follows that $\mathrm{SN}(\rho_B^2) \le 2$. Now consider a unitary channel $\mathcal{F}_B(\rho) = \mathcal{U} \rho_B \mathcal{U}^\dagger$, where the unitary operator $\mathcal{U}$ is given by
\begin{equation}
\mathcal{U} =
\begin{pmatrix}
\frac{1}{\sqrt{3}}+ \frac{1}{\sqrt{6}}& 0 & 0 & 0&0 &0&0&0&-\frac{1}{\sqrt{3}}+ \frac{1}{\sqrt{6}} \\
0 & 1 & 0 & 0 & 0& 0&0& 0& 0 \\
0 & 0 & 1 & 0 &0&0&0&0&0 \\
0 & 0 & 0 & 1 & 0& 0& 0& 0&0\\
-\frac{1}{2\sqrt{3}}+\frac{1}{\sqrt{6}} & 0 & 0 & 0 & \frac{1}{\sqrt{2}}& 0& 0&0& \frac{1}{2\sqrt{3}}+\frac{1}{\sqrt{6}}\\
0 & 0 & 0 & 0 & 0& 1& 0& 0&0\\
0 & 0 & 0 & 0 & 0& 0& 1& 0&0\\
0 & 0 & 0 & 0 & 0& 0& 0& 1&0\\
-\frac{1}{2\sqrt{3}}+\frac{1}{\sqrt{6}} & 0 & 0 & 0 & -\frac{1}{\sqrt{2}}& 0& 0&0& \frac{1}{2\sqrt{3}}+\frac{1}{\sqrt{6}}
\end{pmatrix}.
\end{equation}

Applying this unitary channel onto the state $\rho_B^2$, we obtain the updated state
\begin{equation}
\rho_B^3
= \mathcal{U} \rho_B^2 \mathcal{U}^\dagger
= p \ket{\phi_3^+}\bra{\phi_3^+} + \frac{1-p}{9} I_9.
\end{equation}
where, $\ket{{\phi}^{+}_3} = \frac{1}{\sqrt{3}}(\ket{00} + \ket{11}+\ket{22}) \in \mathbb{C}^3 \otimes \mathbb{C}^3$.
To determine the $2$-Schmidt number of $\rho_B^3$, consider the Schmidt number witness \cite{sanpera2001schmidt}
\begin{equation}
\mathcal{W}_2 = I_9 - \frac{3}{2} \ket{\phi_3^+}\bra{\phi_3^+}.
\end{equation}
A direct calculation shows that
\begin{equation}
\mathrm{Tr}(\mathcal{W}_2 \rho_B^3) < 0 \quad \text{whenever} \quad p > \frac{5}{8}.
\end{equation}
Thus, for $p > \frac{5}{8}$, we have $\mathrm{SN}(\rho_B^3) = 3$. This demonstrates that a non-local channel $\mathcal{F}_B$ can increase the Schmidt number, and hence $\mathcal{F}_B \circ \mathcal{E}$ is not necessarily a non-local $r$-Partially entanglement breaking channel.

   \subsection{Detection of non-local $r$-partially entanglement annihilating channels} In this subsection, we present criteria for detecting channels that are non-local and not $r$-partially entanglement annihilating. We provide two such conditions: the first one is necessary and sufficient, while the remaining one is sufficient condition only.\\

   
  \begin{theorem} \label{theorem4}
       Let $\mathcal{E}$ be a quantum channel acting on subsystem B, where B is a bipartite system. Then   $\mathcal{E} \in$ non-local $r-\mathbb{PEAT}$ iff the map $((id_{B_1} \otimes {\Lambda}) \circ \mathcal{E} )$ is positive, for all $r$ positive but $r+1$ not positive map $\Lambda$.
  \end{theorem}

   \textit{Proof:} Let, ${\mathcal{E}} \in$ non-local $r-\mathbb{PEAT}$. Therefore, SN$({\mathcal{E}} ({\rho}_B)) \le r$ for all  ${\rho}_B \in \mathcal{D}({\mathcal{H}_B})$. Our goal is to show that, the map $((id_{B_1} \otimes {\Lambda}) \circ \mathcal{E} )$ is positive, for all $r$ positive but $r+1$ not positive map $\Lambda$. Now, for arbitrary $\rho_B \in  \mathcal{D}({\mathcal{H}_B})$,
    \begin{equation}
        ((id_{B_1} \otimes {\Lambda}) \circ \mathcal{E} ) (\rho_B) =  (id_{B_1} \otimes {\Lambda})  (\sigma_B) 
    \end{equation}
      where, $ \mathcal{E} ({\rho}_B) = {\sigma}_B$. Since, ${\mathcal{E}} \in$ non-local $r-\mathbb{PEAT}$, therefore  SN$({\sigma}_B) \le r$. Moreover, since ${\Lambda}$ is an $r$ positive map, it preserves positivity when acting on any state with Schmidt number at most $r$ \cite{terhal2000schmidt}. Hence $(id_{B_1} \otimes {\Lambda})  (\sigma_B) \ge 0$. Thus for any ${\rho}_B \in \mathcal{D}({\mathcal{H}_B})$, we get $ ((id_{B_1} \otimes {\Lambda}) \circ \mathcal{E} ) (\rho_B) \ge 0$, which proves that the composed map $((id_{B_1} \otimes {\Lambda}) \circ \mathcal{E} )$ is positive, for all $r$ positive but $r+1$ not positive map $\Lambda$.\\

      Conversely, we want to show that if the map  $((id_{B_1} \otimes {\Lambda}) \circ \mathcal{E} )$ is positive, for all $r$ positive but $r+1$ not positive map $\Lambda$, then $\mathcal{E} \in$ non-local $r-\mathbb{PEAT}$. For arbitrary $\rho_B \in  \mathcal{D}({\mathcal{H}_B})$, consider
      \begin{equation}
        ((id_{B_1} \otimes {\Lambda}) \circ \mathcal{E} ) ( {\rho}_B)  =(id_{B_1} \otimes {\Lambda})  (\sigma_B) \hspace{0.3cm} \text{ where, $ \mathcal{E} ({\rho}_B) = {\sigma}_B$}
            \end{equation}
      By assumption  $((id_{B_1} \otimes {\Lambda}) \circ \mathcal{E} )  ( {\rho}_B)  \ge 0$, which implies that $(id_{B_1} \otimes {\Lambda})  (\sigma_B) \ge 0$, for all $r$ positive but $r+1$ not positive map $\Lambda$.  Hence, SN$(\sigma_B) \le r$  \, \cite{terhal2000schmidt}, which implies   SN$({\mathcal{E}} ({\rho}_B)) \le r$ for all  ${\rho}_B \in \mathcal{D}({\mathcal{H}_B})$ i.e. $\mathcal{E} \in$ non-local $r-\mathbb{PEAT}$.\\

      We, however, note here that this criterion is not an efficient operational method as it requires verification for all $r$-positive but $(r+1)$-not positive maps. Below, we present an alternative condition satisfied by non-local $r$-partially entanglement annihilating channels, with our analysis restricted to covariant channels. Before introducing this sufficient condition, we briefly discuss the notion of absolute Schmidt number channels, which has been introduced recently \cite{mallick2026absolute}.\\

      \textbf{$r$-Absolute Schmidt number channel:} A quantum channel 
\(\mathcal{J}\) acting on subsystem $B$, where $B$ is a bipartite system
is said to be an absolutely \(r\)-Schmidt number channel if, for every bipartite state 
\(\rho_{B} \in \mathcal{D}(\mathcal{H}_{B})\), the Schmidt number of the output state satisfies
\begin{equation}
\mathrm{SN}\!\left( \mathcal{U} \mathcal{J}( \rho_{B} ) \mathcal{U}^{\dagger} \right) \le r.
\end{equation}
for all non-local unitary operators $\mathcal{U} \in \mathbb{U}(d^2)$ where $\mathbb{U}(d^2)$ denotes the set of all non-local unitary operators acting on the composite Hilbert space.

In other words, a quantum channel $\mathcal{S}$ is termed an \(r\) absolutely Schmidt number channel, if under its action, the output state is always belongs to $ r\text{-}\mathbb{ABSN}$, where $ r\text{-}\mathbb{ABSN} = \{\rho \in \mathcal{D} (\mathcal{H}_{B}): \text{SN}(\mathcal{U} \rho {\mathcal{U}}^{\dagger} ) \le r ,\hspace{0.1cm}\forall \hspace{0.1cm}\mathcal{U} \in \mathbb{U} (d^2)\}$. With this definition in place, we now discuss a sufficient condition for a channel to be an $r$-absolutely Schmidt number channel, based on recently developed results on Schmidt number from spectrum \cite{abellanet2026bounding}.\\

\begin{result} \label{result1}
     (Sufficient condition for $r$-absolute Schmidt number channel) \cite{abellanet2026bounding}.
Let $\mathcal{J} : \mathcal{D}(\mathcal{H}_B) \to \mathcal{D}(\mathcal{H}_B)$ be a quantum channel. 
For any input state $\rho_B$, let $\sigma = \mathcal{J}(\rho_B)$ and denote by 
$\lambda = \{\lambda_i^{\uparrow}\}_{i=0}^{d^2-1}$ the eigenvalues of $\sigma$ arranged in non-decreasing order, satisfying $\sum_{i=0}^{d^2-1} \lambda_i = 1$. If, for all input states $\rho_B$, the output spectrum satisfies either
\begin{equation}
\lambda_0(\sigma) \ge \frac{1}{d^2 + \frac{2(dr - 1)}{d - r}},
\end{equation}
or
\begin{equation}
K \sum_{i=0}^{c-1} \lambda_i(\sigma) 
+ \left[ d^2 - Kc + \frac{2(2dr - 1)}{d - r} \right] \lambda_c(\sigma) \ge 1,
\end{equation}
where
\begin{equation}
K = 1 + \frac{2(2dr - 1)}{d - r}, 
\qquad
c = \left\lceil \frac{(d^2 - 1)(d - r)}{2(2dr - 1) - (d - r)} \right\rceil,
\end{equation}
then $\mathcal{J}$ is an absolutely $r$-Schmidt number channel.
\end{result}

With this sufficient criterion for identifying when a channel is an r-absolute Schmidt number, we now use the following result \cite{mallick2026absolute}:

    \begin{result}  \label{result2}\cite{mallick2026absolute} 
       A covariant channel \(\mathcal{J} : \mathcal{D}(\mathcal{H}_{B}) \rightarrow \mathcal{D}(\mathcal{H}_{B})\), is $r$-absolute Schmidt number channel if and only if it is non-local $r$-partially  entanglement annihilating channel.
    \end{result}

Now, combining the above result with Result~\eqref{result1}, we obtain:

\begin{lemma} \label{lemma5}
    For covariant channels, Result \eqref{result1} provides a sufficient condition for a channel to be non-local $r$-partially entanglement annihilating.
\end{lemma}


\subsection{Local partially entanglement annihilating channels}

In this section, we present examples of channels that are not $2$-locally $r$-partially entanglement annihilating. In particular, we consider the qutrit depolarizing and dephasing channels and identify the parameter regimes in which these channels fail to be $2$-locally $r$-partially entanglement annihilating.\ 

\begin{example}
    \textit{(Depolarizing Channel:)} Consider the action of depolarizing channels acting on a $d$-dimensional system defined in Eq.\eqref{depolarizing}. The $2$-local depolarizing channel acts as follows:
\begin{equation}
    \begin{split}
   & \omega_{out} = ({\mathcal{E}} \otimes {\mathcal{E}}) (\omega_{AB}) \\
   & =    p^2 \omega_{AB} + (1-p)^2 (\frac{I_d}{d} \otimes \frac{I_d}{d}) + p(1-p) (\omega_{A} \otimes \frac{I_d}{d} + \frac{I_d}{d} \otimes \omega_{B}) \\ 
   \end{split}
\end{equation}
where, $\omega_A = \text{Tr}_{B}(\omega_{AB})$ and $\omega_B = \text{Tr}_{A}(\omega_{AB})$. Now, to conclude that $ {\mathcal{E}}$ is $2$-locally $r$-partially entanglement annihilating channel, we need to verify for all input states $\omega_{AB}$. Since, the set of states whose Schmidt number less than equal to r $(S_r)$ forms a convex set, therefore it suffices to check only for pure states. Any pure state can be written in the Schmidt decomposition form as :
\begin{equation}
    \ket{\psi} = \sum_j \sqrt{q_j} \ket{\phi_j} \otimes \ket{\widetilde{\phi}_j}
\end{equation}
where $\ket{\phi_j}$ and $\ket{\widetilde{\phi}_j}$ are orthonormal bases on Alice's and Bob's side respectively. Now, $\omega_{AB} = \ket{\psi}\bra{\psi}$ and the reduced states take the form 
\begin{equation}
  \omega_A  = \sum_j q_j \ket{\phi_j} \bra{\phi_j} \hspace{0.2cm} \text{and}  \hspace{0.2cm} \omega_B  = \sum_j q_j \ket{\widetilde{\phi}_j} \bra{\widetilde{\phi}_j} 
\end{equation}
In \cite{moravvcikova2010entanglement}, authors have considered qubit depolarizing to analyze 2-locally entanglement annihilating channel. They have shown that, the qubit depolarizing channel is 2-locally entanglement annihilating channel iff  $p \le \frac{1}{\sqrt{3}}$. Also, it is known that \cite{horodecki2003entanglement,moravvcikova2010entanglement} qubit depolarizing channel is entanglement breaking for $p \le \frac{1}{{3}}$. Since partially entanglement breaking and annihilating channel is more general than entanglement breaking and annihilating channel, therefore here we will consider qutrit $(d=3)$ depolarizing channel to check whether it is 2-locally partially entanglement annihilating channel. Now $\ket{\psi}$  takes the form 
\begin{equation}
    \ket{\psi} =  \sqrt{q_0} \ket{\phi_0} \otimes \ket{\widetilde{\phi}_0} +  \sqrt{q_1} \ket{\phi_1} \otimes \ket{\widetilde{\phi}_1} +  \sqrt{q_2} \ket{\phi_2} \otimes \ket{\widetilde{\phi}_2}
\end{equation}

Applying  ${\mathcal{E}} \otimes {\mathcal{E}}$ on $\omega_{AB}$, we get
\begin{equation} 
\begin{split}     
& \omega_{out}= \begin{bmatrix} 
	\frac{(p^2+2p)q_0}{3}+ t &0& 0 &0 &p^2 \sqrt{q_0 q_1}&0 &0&0&p^2 \sqrt{q_0 q_2}\\[0.1cm]
    0& 	s_1+t &0&0&0&0&0&0 &0\\[0.1cm]
    0&0& 	s_2+t	&0&0&0&0&0&0\\[0.1cm]
     0&0& 0	&s_1+t &0&0&0&0&0\\[0.1cm]
     p^2 \sqrt{q_0 q_1}&0&0& 0&\frac{(p^2+2p)q_1}{3}+ t &0&0&0&p^2 \sqrt{q_1 q_2}\\[0.1cm]
    	 0 &0&0&0&	0&s_3+t &0&0&0 \\[0.1cm]
    0& 0&0&0&0&0&s_2+t&0&0\\[0.1cm]
    0& 0&0&0&0&0&0&s_3+t&0\\[0.1cm]
  p^2 \sqrt{q_0 q_2}&0&0&0&p^2 \sqrt{q_2 q_1}&0&0&0& \frac{(p^2+2p)q_2}{3}+ t\\[0.1cm] \nonumber
	\end{bmatrix}
 \end{split}
	\end{equation}
where, $t=\frac{(1-p)^2}{9}$, $s_1=\frac{{p(1-p)}(q_0 +q_1)}{3}$, $s_2=\frac{{p(1-p)}(q_0 +q_2)}{3}$ and $s_3=\frac{{p(1-p)}(q_1 +q_2)}{3}$.\\

 Now, we know that \cite{terhal2000schmidt} the set of $r$ positive but $r+1$ not positive maps acts as a witness for $S_r$. Since, any  $r$ positive but $r+1$ not positive map  $({\Lambda}_k)$ should satisfy  $(id_A \otimes {\Lambda}_k )(\rho) \ge 0$ for all  $\rho \in S_r$ and $(id_A \otimes {\Lambda}_k )(\sigma) < 0$ for at least one  $\sigma \in S_{r+1}$, where $S_r$ denotes the set of states whose Schmidt number is less than or equal to $r$. An example of such ${\Lambda}_k$ defined on ${\mathcal{M}}_d$ be of the form
\begin{equation}\label{r-reductionmap}
    {\Lambda}_k (X) = \text{Tr}(X) I_d - kX \hspace{.2cm} \text{for} \hspace{.2cm} X \in {\mathcal{M}}_d   
\end{equation}
 The map ${\Lambda}_k (X)$ is $r$ positive but $r+1$ not positive map (where, $r <d $) for 
\begin{equation} \label{rpositive}
    \frac{1}{r+1} < k \le \frac{1}{r} 
\end{equation} 

Since the map ${\Lambda}_k$ is not sufficient to conclude for which $p$ the channel $\mathcal{E}$ is  $2$-locally partially entanglement annihilating channel. But if we get $(id_A \otimes {\Lambda}_k )( \omega_{out}) < 0$ for a map that is $2$-positive but $3$ not positive, then we can safely say that the channel $\mathcal{E}$ is not a 2-locally partially entanglement annihilating channel. Now, From Eq.\eqref{rpositive}, we determine that the map ${\Lambda}_k$ is $2$-positive but $3$ not positive map for $ \frac{1}{3} < k \le \frac{1}{2}$. For simplicity here we take $k= \frac{1}{2}$. We obtain the minimum eigen value of $(id_A \otimes {\Lambda}_k )( \omega_{out})$ occurs for $ q_0 = q_1=q_2= \frac{1}{3}$ and the minimum eigen value is $\frac{2-8p^2}{9}$. It can be checked easily that for $p>\frac{1}{2}$, the eigenvalue $\frac{2-8p^2}{9}$ is not positive. Therefore, the qutrit depolarizing channel is not a 2-locally partially entanglement annihilating channel for $p > \frac{1}{2}$.
\end{example} 
\begin{example}
    \textit{(Dephasing channel:)} Consider the action of a qutrit dephasing channel, as given in Eq.~\eqref{dephasing}. Following a procedure analogous to the previous example, the action of the $2$-local qutrit dephasing channel on an arbitrary bipartite state $\omega_{AB}$ yields the output state as 
\begin{equation} 
\begin{split}
& \omega_{out} =(\mathcal{D} \otimes \mathcal{D})(\omega_{AB})=\begin{bmatrix} 
	q_0 &0& 0 &0 &v^2 \sqrt{q_0 q_1}&0 &0&0&v^2 \sqrt{q_0 q_2}\\[0.1cm]
    0& 	0 &0&0&0&0&0&0 &0\\[0.1cm]
    0&0& 	0	&0&0&0&0&0&0\\[0.1cm]
     0&0& 0	&0 &0&0&0&0&0\\[0.1cm]
     v^2 \sqrt{q_0 q_1}&0&0& 0&q_1 &0&0&0&v^2 \sqrt{q_1 q_2}\\[0.1cm]
    	 0 &0&0&0&	0&0 &0&0&0 \\[0.1cm]
    0& 0&0&0&0&0&0&0&0\\[0.1cm]
    0& 0&0&0&0&0&0&0&0\\[0.1cm]
  v^2 \sqrt{q_0 q_2}&0&0&0&v^2 \sqrt{q_2 q_1}&0&0&0& q_2\\[0.1cm] \nonumber
	\end{bmatrix}
 \end{split}
	\end{equation}
After applying the $2$-positive but not $3$-positive map $\Lambda_k$ defined in Eq.~\eqref{r-reductionmap}, with $k=\tfrac{1}{2}$, the minimum eigenvalue of the resulting output state is given by $\tfrac{1}{2} - v^2$ and this minimum is attained for the uniform choice $q_0 = q_1 = q_2 = \tfrac{1}{3}$. Consequently, the qutrit depolarizing channel fails to be a $2$-locally partially entanglement annihilating channel whenever $v > \tfrac{1}{\sqrt{2}}$.
\end{example}



\section{Relations between $\mathbb{EB}$, $r-\mathbb{PEBT}$, $2$-locally $r-\mathbb{PEAT}$ and non-local $r-\mathbb{PEAT}$} \label{s6}
In this section, we explore the relationships between  $\mathbb{EB}$, $r-\mathbb{PEBT}$ and $r-\mathbb{PEAT}$. We begin by examining the connection between $\mathbb{EB}$ and $r-\mathbb{PEBT}$. Subsequently, we discuss the relationships between $r-\mathbb{PEBT}$ and $2$-locally $r-\mathbb{PEAT}$, as well as non-local $r-\mathbb{PEAT}$ and $r-\mathbb{PEBT}$, all defined on the same Hilbert space  ${\mathcal{H}}_B$. \\

\textbf{Relation between $\mathbb{EB}$ and $r-\mathbb{PEBT}$ :} 
We compare $r-\mathbb{PEBT}$ and $\mathbb{EB}$ by examining depolarizing channel, as defined in Eq. \eqref{depolarizing}. Specifically, for a depolarizing channel \cite{chruscinski2006partially}:
\begin{itemize}
    \item The channel is entanglement breaking if $p \le \frac{1}{d+1}$
    \item The channel is r-partially entanglement breaking channel if and only if  $ p \le \frac{r d-1}{d^2-1} $
\end{itemize}
Thus, while all entanglement breaking channels are r-partially entanglement breaking channels, the converse is not true i.e. $\mathbb{EB}$ is a strict subset of $r-\mathbb{PEBT}$. This distinction is evident in cases where  $r < d$ and $p$ lies in the range
\begin{equation}
  \frac{1}{d+1} < p \le  \frac{r d-1}{d^2-1}
\end{equation}
In this interval, the depolarizing channel meets the r-partially entanglement breaking channel criterion without being entanglement breaking, indicating that the set $r-\mathbb{PEBT}$ is larger than the set of $\mathbb{EB}$. Consequently, the following inclusion relation holds among these sets:
\begin{equation}
    \text{$\mathbb{EB}$} = \text{$1-\mathbb{PEBT}$} \subseteq \text{$2-\mathbb{PEBT}$} \subseteq \text{$3-\mathbb{PEBT}$} ......  \subseteq  \text{$r-\mathbb{PEBT}$}
\end{equation}

\textbf{Relation between $r-\mathbb{PEBT}$ and $2$-locally $r-\mathbb{PEAT}$ :} To analyze the relationship between $r-\mathbb{PEBT}$ and $2$-locally $r-\mathbb{PEAT}$,  let us first assume that $\mathcal{R} \in$  $r-\mathbb{PEBT}$. Now, consider the composite map $\mathcal{R} \otimes \mathcal{R}$. We can express its action on an arbitrary density operator $\sigma_B \in \mathcal{D(H_B)}$ can be written as:
\begin{equation}
     ( \mathcal{R} \otimes \mathcal{R} ) ({\sigma}_B)  
    =   ( \mathcal{R} \otimes id_B) (id_A \otimes \mathcal{R} ({\sigma}_B) \Big)
    =  ( \mathcal{R} \otimes id_B) ({\sigma}^1_B) \Big)
    = {\sigma}^2_B
 \end{equation}
where, $B$ is a bipartite system and ${\sigma}^1_B = (id_A \otimes \mathcal{R}) (\sigma_B)$. Since, $\mathcal{R} \in$  $r-\mathbb{PEBT}$, therefore SN$({\sigma}^2_B) \le r$. Hence, $\mathcal{R} \in$ 2-local  $r-\mathbb{PEAT}$. It follows that, every r-partially entanglement breaking channel is necessarily a $2$-locally  r-partially entanglement annihilating channel i.e. 
\begin{equation}
   \text{ $r-\mathbb{PEBT}$} \subset \text{ 2-local $r-\mathbb{PEAT}$}
\end{equation} \\
Now, we will discuss whether the converse relation between $r-\mathbb{PEBT}$ and 2-local $r-\mathbb{PEAT}$ holds; specifically, whether 2-local  $r-\mathbb{PEAT}$ is a subset of  $r-\mathbb{PEBT}$. For this purpose, let us first consider the qubit depolarizing channel,
\begin{equation} \label{qubitdepolarizing}
    {\mathcal{E}}_2 (\rho) = p \rho + (1-p)  \text{Tr}(\rho) \hspace{0.1cm} \frac{I_2}{2}
\end{equation}
where $p \in [0,1]$. It is known that \cite{moravvcikova2010entanglement}, this channel belongs to 2-locally $\mathbb{EA}$ iff $p \le \frac{1}{\sqrt{3}}$ and  $\mathbb{EB}$ iff $p \le \frac{1}{{3}}$. Therefore, for the range $p \le \frac{1}{\sqrt{3}}$ the channel belongs to 2-locally $\mathbb{EA}$ as well as $\mathbb{EB}$ but for the range $\frac{1}{{3}} < p \le \frac{1}{\sqrt{3}}$, the qubit depolarizing channel belongs to 2-locally $\mathbb{EA}$ but not in $\mathbb{EB}$. This shows that 2-locally $\mathbb{EA}$ is not a subset of $\mathbb{EB}$. Since $\mathbb{EB}$ and $\mathbb{EA}$ are the special cases of $r-\mathbb{PEBT}$ and $r-\mathbb{PEAT}$ respectively when $r=1$. Therefore, there exists at least one 2-locally $r$-partially entanglement annihilating channel which is not an $r$-partially entanglement breaking channel. Hence, we can conclude that 
\begin{equation}
     \text{2-local $r-\mathbb{PEAT}$} \not\subset \text{$r-\mathbb{PEBT}$}
\end{equation}



 \textbf{Relation between $r-\mathbb{PEBT}$ and non-local $r-\mathbb{PEAT}$ :} Here, we will discuss the set relation between $r-\mathbb{PEBT}$ and non-local  $r-\mathbb{PEAT}$. Let $\mathcal{S}$ and $\mathcal{E}$ be an r-partially entanglement breaking channel and non-local r-partially entanglement annihilating channel defined on the same Hilbert space $H_B$. From the definition of r-partially entanglement breaking channel and non-local r-partially entanglement annihilating channel, it is obvious that the composition of these two channels i.e. $\mathcal{S} \circ \mathcal{E}$ always belongs to $r-\mathbb{PEBT}$, as well as non-local $r-\mathbb{PEAT}$. Therefore, there exists at least one channel which belongs to both $r-\mathbb{PEBT}$ and non-local $r-\mathbb{PEAT}$. Hence, 
 \begin{equation}
    \text{$r-\mathbb{PEBT}$} \hspace{0.1cm}\cap  \hspace{0.1cm} \text{non-local $r-\mathbb{PEAT}$} \neq \emptyset
 \end{equation}

 Now, let us assume that $r-\mathbb{PEBT}$ $\subset$ non-local $r-\mathbb{PEAT}$. Let,  $\mathcal{S} \in$ $r-\mathbb{PEBT}$. Since,  $r-\mathbb{PEBT}$ $\subset$ non-local $r-\mathbb{PEAT}$, therefore  $\mathcal{S} \in$ non-local $r-\mathbb{PEAT}$. we take ${\mathcal{F}}_B$ be any quantum channel acting on subsystem $B$. From Proposition 1 and Proposition 3, we deduce that   $\mathcal{S} \circ {\mathcal{F}}_B $ always belongs to $r-\mathbb{PEBT}$ but ${\mathcal{F}}_B \circ \mathcal{S} $ may not be non-local $r-\mathbb{PEAT}$, since  Schmidt number of a state can increase under the action of ${\mathcal{F}}_B$,  For example, if ${\mathcal{F}}_B$ is a non-local unitary operation, then the Schmidt number of a state could increase. Therefore, there exists at least one channel that belongs to $r-\mathbb{PEBT}$ but does not belong to non-local $r-\mathbb{PEAT}$. This contradicts our assumption. Hence,
 \begin{equation}
     \text{$r-\mathbb{PEBT}$} \hspace{0.1cm} \not\subset \hspace{0.1cm} \text{non-local $r-\mathbb{PEAT}$}
 \end{equation}

   \section{Conclusions} \label{s7}

The dimensionality of entanglement is characterized by Schmidt numbers. Higher Schmidt numbers carry signatures of high dimensional entanglement.    
They act as a valuable resource, offering enhanced performance in various quantum information processing tasks \cite{bae2019more,cozzolino2019high,kues2017chip,wang2020qudits}. Consequently, quantum channels that reduce the Schmidt number of such states are termed partially entanglement breaking channels \cite{chruscinski2006partially}. In this work, we have characterized the partially entanglement breaking channels and discussed their properties and behavior. We have proved that the set of all $r$ partially entanglement breaking channel is a convex and compact set. This allows one to detect non-partially entanglement breaking channels. In particular, we have determined the exact parameter ranges in which the depolarizing and dephasing channels do not act as partially entanglement breaking channels. Moreover, we obtain the sufficient condition to detect non-r-
partially entanglement breaking channels. Furthermore, we have shown that the series concatenation of two partially entanglement breaking channels results in another partially entanglement breaking channel. This property is not preserved under parallel concatenation. 

We then introduce the notion of partially entanglement annihilating channels and provide their topological characterization. By considering the qutrit depolarizing and dephasing channels, we provide the parameter regime in which this channel acts as a non-partially entanglement annihilating channel. A sufficient condition for r-partially entanglement annihilating channels has been also derived. From the perspective of resource theory of channels, it is imperative to identify channels which not break a resource, so that they can be used reliably in an information processing protocol. The characterization done in this work facilitates this identification. Finally, we examine the relationships between entanglement breaking, partially entanglement breaking, and partially entanglement annihilating channels.
Our study reveals various interesting connections among the sets consisting
of the above categories of channels. 

Our study opens several avenues for future research. While we have provided a topological characterization of partially entanglement annihilating channels, obtaining a Choi-Kraus type representation for such channels remains an important open problem.  We have also shown that $r$-partially entanglement breaking channels are not closed under tensor products, motivating the investigation of whether a similar property holds for $r$-partially entanglement annihilating channels. The study of communication and entanglement transmission capacities of partially entanglement breaking and annihilating channels deserves further attention. Recent work has highlighted the operational significance of $r$-partially entanglement breaking channels by showing that any such channel can be implemented via one-way LOCC assisted by an entangled state of Schmidt number $r$ \cite{namiki2013composability}. The characterization of these channels through a general Holevo-type form, analogous to entanglement breaking channels \cite{horodecki2003entanglement}, remains an interesting open direction. Finally, extending the notions of partially entanglement breaking and annihilating channels to multipartite systems constitutes a promising line of investigation due to the rich and inequivalent structures of multipartite entanglement and entanglement dimensionality \cite{huber2013structure}.

\begin{acknowledgments}
B.M. acknowledges Sahil Gopalkrishna Naik for fruitful discussions. B.M. also acknowledges the DST INSPIRE fellowship program for financial support. N.G. acknowledges support from DST-SERB MATRICS grant vide file number MTR/2022/000101.
\end{acknowledgments}

\section*{Data Availability Statement}
Data sharing is not applicable to this article as no new data were created or analyzed in this study.

\bibliography{main}
\appendix 

\section{Proofs of Lemma \eqref{lemma3} and Lemma \eqref{lemma4}} \label{appendixA}
\textbf{Proof of Lemma \eqref{lemma3}:} Let, ${\mathcal{E}}_1$, ${\mathcal{E}}_2 \in$ non-local $r-\mathbb{PEAT}$. Therefore, 
\begin{equation}
    \text{SN} ({\mathcal{E}}_1 ({\rho}_B)) \le r \hspace{0.5cm} \text{and} \hspace{0.5cm} \text{SN} ({\mathcal{E}}_2 ({\rho}_B)) \le r
\end{equation} for all ${\rho}_B \in \mathcal{D}({\mathcal{H}_B})$. Our aim is to show that for any $p \in [0,1]$, the convex combination $ \mathcal{E} =p \mathcal{E}_{1} + (1-p) \mathcal{E}_{2} $, also belongs to non-local $r-\mathbb{PEAT}$.
Now, the action of the channel $ p \mathcal{E}_{1} + (1-p) \mathcal{E}_{2} $ on an arbitrary density operator $\sigma_B \in \mathcal{D}({\mathcal{H}_B})$ can be written as \begin{equation}
     \mathcal{E} (\sigma_B)  
    =    \Big(  (p \mathcal{E}_{1} + (1-p) \mathcal{E}_{2})  (\sigma_B) \Big)
   =   \Big(  p  {\sigma}^{1}_B + (1-p) {\sigma}^{2}_B   \Big) \hspace{0.5cm} 
\end{equation}
where, ${\sigma}^{1}_B = \mathcal{E}_{1} (\sigma_B)$ and ${\sigma}^{2}_B = \mathcal{E}_{2}    (\sigma_B)$. Since, ${\mathcal{E}}_1$, ${\mathcal{E}}_2 \in$ non-local $r-\mathbb{PEAT}$, therefore SN$({\sigma}^{1}_B)$ and SN$({\sigma}^{2}_B)$ are less than equal to $r$. It is known that the set of states whose Schmidt number is less than or equal to $r$ forms a convex set \cite{terhal2000schmidt}. Hence, SN$( p  {\sigma}^{1}_B + (1-p) {\sigma}^{2}_B)$ is also less than or equal to $r$. Thus we conclude that SN$(\mathcal{E} (\sigma_B)) \le r$, which proves that non-local $r-\mathbb{PEAT}$ is convex.\\

\textbf{Proof of Lemma \eqref{lemma4}:} First, we will prove that the set, non-local $r-\mathbb{PEAT}$ is closed, i.e., it contains all of its limit points. Let $\mathcal{E}_{0}$ be an arbitrary limit point of non-local $r-\mathbb{PEAT}$ ( since the set is convex, therefore, it always has a limit point), therefore, if we consider an open ball $B_{\frac{1}{m}}(\mathcal{E}_{0})$ of radius $\frac{1}{m}$ centered on $\mathcal{E}_{0}$, then 
\begin{equation}
   \{ B_{\frac{1}{m}}(\mathcal{E}_{0}) - \mathcal{E}_{0} \} \cap \text{non-local $r-\mathbb{PEAT}$} \neq \emptyset
\end{equation}
Since, each neighborhood of $\mathcal{E}_{0}$  contains infinitely many points of non-local $r-\mathbb{PEAT}$, where $\emptyset$ indicates the null set.
Let us now construct a sequence $\{\mathcal{E}_m\}$ of distinct non-local $r$-partially entanglement annihilating channels such that   $\mathcal{E}_m \rightarrow \mathcal{E}_{0} $ as follows:
\begin{align}
    & \mathcal{E}_1 \in B_1(\mathcal{E}_{0}) \cap \text{non-local $r-\mathbb{PEAT}$} , \hspace{0.15cm} \mathcal{E}_1 \neq \mathcal{E}_{0}\\ \nonumber
     &  \mathcal{E}_2 \in  B_{\frac{1}{2}}(\mathcal{E}_{0}) \cap \text{non-local $r-\mathbb{PEAT}$} ,  \hspace{0.15cm} \mathcal{E}_2 \neq \mathcal{E}_{0},  \mathcal{E}_1 \\ \nonumber
      &   .... \in ....  \\ \nonumber
       &   .... \in ....  \\ \nonumber
         &  \mathcal{E}_m \in  B_{\frac{1}{m}}(\mathcal{E}_{0}) \cap \text{non-local $r-\mathbb{PEAT}$} , \hspace{0.15cm}  \mathcal{E}_m \neq  \mathcal{E}_{0},...,\mathcal{E}_{m-1}
   \end{align}
From the above construction, it is evident that $\mathcal{E}_m \rightarrow \mathcal{E}_{0}$. Now, we take another sequence $\{ \mathcal{E}_m ({\rho}_B) = {\chi^B_m} \}$ for arbitrary density operator ${\rho}_B \in  \mathcal{D}({\mathcal{H}_B})$. Since, $\mathcal{E}_m \in $ non-local $r-\mathbb{PEAT}$, therefore, 
\begin{equation}
    \text{SN} [\mathcal{E}_m ({\rho}_B)] = \text{SN} (\chi^B_m) \leq r
\end{equation}
We know that the set of states $(S_r)$ whose Schmidt number is less than equal to $r$ forms a closed set \cite{terhal2000schmidt}, therefore for $ \chi^B_m \in S_r $, all limit points of $\chi^B_m$ also belongs to $ S_r$. Now, 
\begin{align}
   & \lim_{m\to\infty}  \mathcal{E}_m ({\rho}_B) \rightarrow  \mathcal{E}_0 ({\rho}_B)  \\ \nonumber
   & \implies \lim_{m\to\infty} \chi^B_m = \chi^B_0   \hspace{1cm} [\text{where,} \hspace{0.2cm}  \mathcal{E}_0 ({\rho}_B) = {\chi^B_0}]
\end{align}
Therefore, 
$\chi^B_0 \in  S_r$, which implies 
\begin{equation}
    \text{SN} [\mathcal{E}_0 ({\rho}_B)] \leq r, \hspace{0.3cm} \text{for arbitrary ${\rho}_B \in  \mathcal{D}({{\mathbf{C}}^d \otimes  {\mathbf{C}}^d})$}
\end{equation}
i.e. $\mathcal{E}_0 \in $ non-local $r-\mathbb{PEAT}$. Since we have considered $\mathcal{E}_0$ to be an arbitrary limit point of non-local $r-\mathbb{PEAT}$, it follows that non-local $r-\mathbb{PEAT}$ contains all of its limit points. Therefore, the set non-local $r-\mathbb{PEAT}$ is closed.

Again, we know that the completely bounded trace norm of
quantum channels is equal to 1, \cite{watrous2018theory} therefore the channels in the set non-local $r-\mathbb{PEAT}$ are bounded maps. Hence, the set non-local $r-\mathbb{PEAT}$ is both closed and bounded, i.e. compact.\\



%

\end{document}